\documentclass[12pt]{article} 
\usepackage{axodraw}
\usepackage{amsmath,amsfonts,amssymb}
\usepackage{color}
\newcommand{\be}{\begin{equation}}
\newcommand{\ee}{\end{equation}}
\newcommand{\bear}{\begin{eqnarray}}
\newcommand{\eear}{\end{eqnarray}}
\newcommand{\ba}{\begin{array}}
\newcommand{\ea}{\end{array}}

\begin{document}

\baselineskip=18pt \pagestyle{plain} \setcounter{page}{1}

\vspace*{-1cm}

\noindent 
\makebox[12.3cm][l]{\small \hspace*{-.2cm}  August 17, 2015; revised December 1, 2015}{\small Fermilab-Pub-15-349-T}  \\  [-1mm]
\makebox[12.8cm][l]{\small \hspace*{-.2cm} }{\small SISSA 36/2015 FISI} \\ [-2mm]

\begin{center}

{\Large \bf  Right-handed neutrinos and the 2 TeV $W'$ boson   } \\ [9mm]

{\normalsize \bf Pilar Coloma$^\star$, Bogdan A. Dobrescu$^\star$, Jacobo Lopez-Pavon$^\diamond$ \\ [5mm]
{\small {\it
$\star$ Theoretical Physics Department, Fermilab, Batavia, IL 60510, USA   \\ [2mm]
 $\diamond$ \parbox[t]{11.5cm}{ SISSA and INFN sezione di Trieste, 34136 Trieste, Italy \\
 {\rm  \&} INFN, Sezione di Genova, via Dodecaneso 33, 16146 Genova, Italy}
 }}\\
}

\end{center}

\vspace*{0.2cm}

\begin{abstract}
The CMS $e^+e^-jj$ events of invariant mass near 2 TeV are consistent with a $W'$ boson decaying into an electron and a 
right-handed neutrino whose TeV-scale mass is of the Dirac type. We show that the Dirac partner of the right-handed electron-neutrino
can be the right-handed tau-neutrino. 
A prediction of this model is that the sum of the  $\tau^+ e^+  jj$ and $\tau^- e^-  jj$  signal cross sections equals twice 
that for $e^+e^-jj$. The Standard Model neutrinos acquire Majorana masses 
and mixings compatible with neutrino oscillation data. 
\end{abstract}


\vspace*{0.5cm}

\section{Introduction} \setcounter{equation}{0}

Searches for new gauge bosons that carry electric charge $\pm 1$, called $W'$ bosons, have been intensely performed 
at hadron colliders \cite{Agashe:2014kda}. More than a year ago, 
the CMS Collaboration \cite{Khachatryan:2014dka} has reported an excess of events in the search for a $W'$ boson
decaying into a heavy right-handed neutrino and an electron. The observed final state 
consists of an electron-positron pair and two hadronic jets, with an invariant mass in the 1.8--2.2 TeV range. 
Although this excess has a statistical significance of only 2.8$\sigma$, it is particularly interesting given the
theory connection \cite{Dobrescu:2015qna} based on the $SU(2)_R$ gauge group between this and 
other excess events observed at the LHC in the $W' \to jj$  \cite{Khachatryan:2015sja}, $W' \to WZ$ \cite{Aad:2015owa,Khachatryan:2014gha}  
and $W' \to Wh^0$  \cite{CMS:2015gla} channels.

A key feature of the CMS $e^+ e^-jj$ excess is that it is not accompanied by an excess in the $e^-e^-jj $ and $e^+e^+jj $
final states. In the traditional $SU(2)_L\times SU(2)_R\times U(1)_{B-L}$ model, where the right-handed
fermions form $SU(2)_R$ doublets \cite{Mohapatra:1986uf}, the right-handed neutrinos have Majorana masses. As a consequence, 
the cross section for the final state with same-sign leptons ($e^\mp e^\mp jj$) is equal to that for opposite-sign leptons  ($e^+ e^-jj$) \cite{Keung:1983uu,Han:2012vk}.
Given this generic prediction and the smaller backgrounds  for same-sign leptons, the ATLAS searches \cite{Aad:2015xaa} have 
not included $e^+ e^- jj$ resonances. 

An explanation for the lack of a same-sign signal has been proposed in  \cite{Dobrescu:2015qna}, where the right-handed neutrino $N_R$ partners with 
the neutral component of an $SU(2)_R$ doublet fermion and acquires a Dirac mass. This solution works with the minimal Higgs sector \cite{Dobrescu:2015yba}, 
namely a bidoublet and an $SU(2)_R$ triplet. A related solution, proposed in 
 \cite{Dev:2015pga}, is to generate Dirac masses for right-handed neutrinos by introducing three gauge-singlet fermions while extending the Higgs sector with 
 an $SU(2)_R$-doublet scalar. 
 Both these solutions also include subdominant  Majorana masses for right-handed neutrinos, so that the physical states are actually pseudo-Dirac. 
 Given that only one of the 14 CMS $eejj$ events has same-sign leptons, which may be due to the background (estimated to be approximately four $eejj$ events),
 it is safe to neglect the Majorana component.

On the other hand it has been pointed out  in \cite{Gluza:2015goa} that particular values for the Majorana masses of the right-handed neutrinos 
and their CP-violating phases may allow a suppression of the 
same-sign $eejj$ signal, without introducing additional fields.
This may be counter-intuitive, given that the  $W' \!\to eN$ decays produce very narrow on-shell $N$ particles, whose decay widths
should not be sensitive to the CP violating phases. The suppression is provided by interference effects between processes proceeding through different right-handed 
neutrinos. Other studies of the $W'$ interpretation of the CMS $eejj$ excess can be found in \cite{Krauss:2015nba}.

Here we propose that the Dirac partner of the right-handed electron-neutrino is the 
right-handed $\tau$-neutrino. 
We show that the flavor structure required by this mechanism can easily be enforced by a symmetry.
The ensuing model is remarkably simple
and leads to peculiar signals at the LHC.  This mechanism also explains the suppression of same-sign $eejj$ signals observed in the simulations 
discussed in \cite{Gluza:2015goa}: 
our Dirac fermion can be decomposed in two degenerate Majorana fermions whose interactions with the $W'$ boson include imaginary couplings.  

In Section 2 we show how a Dirac state arises from two right-handed neutrino flavors, and derive its interactions.
The implications for LHC phenomenology are analyzed in Section 3. Mechanisms for generating masses for the Standard Model (SM) neutrinos in this context are 
discussed in Section 4. We summarize our conclusions in Section 5.

\bigskip 

\section{Two right-handed neutrinos make a Dirac fermion}\setcounter{equation}{0}

We consider an $SU(3)_c \times SU(2)_L \times SU(2)_R \times U(1)_{B-L}$ gauge theory with a minimal 
 fermion content: the three generations of SM quarks and leptons plus one right-handed neutrino per generation. 
The three right-handed neutrinos 
($N_R^e$, $N_R^\mu$, $N_R^\tau$) together with the three right-handed charged leptons ($e_R$, $\mu_R$, $\tau_R$) form 
$SU(2)_R$ doublets of $U(1)_{B-L}$ charge $-1$. We label these by $L^e_R =(N_R^e, e_R)^\top   $, 
$L^\mu_R =(N_R^\mu, \mu_R)^\top$ and $L^\tau_R =(N_R^\tau, \tau_R)^\top   $. 

The minimal Higgs sector consistent with the $W'$ signals discussed in the Introduction has been analyzed in detail in 
Ref.~\cite{Dobrescu:2015yba}. It includes an $SU(2)_R$ triplet scalar $T$ of  $U(1)_{B-L}$ charge +2, and 
an $SU(2)_L \times SU(2)_R$ bidoublet scalar $\Sigma$ that is not charged under $U(1)_{B-L}$.
The VEV of $T$ breaks $SU(2)_R \times U(1)_{B-L}$ down to the hypercharge gauge group, $U(1)_Y$, while inducing 
the required mass splitting between $W'$ and $Z'$ \cite{Dobrescu:2015qna}. 

The VEV of $\Sigma$ induces a mass mixing between the $SU(2)_L \times SU(2)_R$ gauge bosons
that carry electric charge  
($W_L$ and $W_R$), so that the physical spin-1 particles ($W$ and $W'$) couple together to the $Z$ boson.
This VEV breaks the electroweak symmetry, and can be parametrized as
$\langle \Sigma \rangle = v_H\,  {\rm diag} \, (\cos\beta , e^{i\alpha_\Sigma} \sin\beta) $, where $v_H \simeq 174$ GeV.
The excess events in the $W'\to WZ$ searches \cite{Aad:2015owa,Khachatryan:2014gha} indicate a $W'WZ$ coupling 
consistent with near-maximal $W_L-W_R$ mixing, which corresponds to $\tan \beta \approx 1$.

The right-handed lepton doublets have gauge-invariant Yukawa couplings to the triplet scalar. 
Our main assumption is that these Yukawa couplings have the following flavor structure:
\be
-  \frac{y_{\mu\mu}}{2}  (\, \overline L^\mu_R)^c \; i \sigma_2 \, T \, L^\mu_{R}
- y_{e\tau}   (\, \overline L^e_{R})^c \; i \sigma_2 \, T \, L^\tau_{R}
+ {\rm H.c.}   ~~,
\label{eq:flavorYuk}
\ee
where $y_{\mu\mu}$ and $y_{e\tau}$ are positive dimensionless parameters, 
and $\sigma_2$ is the Pauli matrix acting on $SU(2)_R$ representations. The $c$ 
label here indicates, as usual,
the charge-conjugate spinor \cite{Cheng:1985bj}. 
We will show in Section 3 that this non-diagonal structure in flavor space implies the absence of same-sign $eejj$ events at the LHC.

The flavor structure of the Yukawa couplings in Eq.~(\ref{eq:flavorYuk}) can be enforced by a symmetry. An example is a a global $U(1)$ symmetry
with the  $L^e_R$,  $L^\mu_{R}$,  $L^\tau_{R}$ doublets carrying charges  $-1,0,+1$, respectively; the scalar $T$ must then be neutral under this 
global symmetry. Even a discrete subgroup ($Z_n$ with $ n \ge 3$)
of this $U(1)$ symmetry would be sufficient.  \\

Once the scalar triplet $T$ gets a VEV,
\be 
\langle T \rangle = \left( \ba{cc}   0 & 0  \\ u_T & 0   \ea \right)  ~,
\ee
the right-handed neutrinos acquire masses through the following Lagrangian terms:
\be
\label{MR}
- u_T \left(  \, \overline{N}_R^e ,  \, \overline N_R^\mu,  \, \overline N_R^{\, \tau} \right)^c
 \left( \ba{ccc}   0 & 0 & y_{e\tau}  \\ 0 & y_{\mu\mu}  & 0  \\   y_{e\tau}  & 0 & 0  \ea \right)  
 \left( \ba{c}     N_R^e \\ [1mm] N_R^\mu \\ [1mm] N_R^\tau    \ea \right)  ~~.   
 \ee
The parameter $u_T$ is in the 3--4 TeV range \cite{Dobrescu:2015yba} in order to accommodate the mass and coupling of the $W'$ boson 
indicated by the LHC data.

The 2-component fermion  $N_R^\mu$ is already in the mass eigenstate basis, and has a Majorana 
mass $m_{N^\mu} = y_{\mu\mu} u_T$. 
More importantly, we find that $N_R^e$ and $N_R^\tau$ form a 4-component fermion $N_1$
of Dirac mass
\be
m_{N_1} =  y_{e\tau} u_T  ~~. 
\ee

The interactions of the $W'$ boson with the right-handed neutrinos in the gauge eigenstate basis
are given by
\be
\frac{g_{_{\rm R}} } {\sqrt 2}  \, W'_\nu \, \left(  \overline N_R^e \gamma^\nu e_R +  \overline N_R^\mu \gamma^\nu \mu_R  
+ \overline N_R^\tau \gamma^\nu \tau_R \right)  + {\rm H.c.} ~~,
\label{eq:WpInt}
\ee
where the $g_{_{\rm R}}$ coupling is equal to the $SU(2)_R$ gauge coupling up to negligible corrections. 
We identify the $N_R^\tau$ and $N_R^e$ fields with the left- and right-handed components of the Dirac fermion $N_1$:
\be
N_R^\tau \equiv N_{1_L}^c       \;\;\;\;   ,    \;\;\;\;    N_R^e \equiv N_{1_R}     ~~~.
\ee
The spinors satisfy the usual  \cite{Cheng:1985bj}  relation  $N_{1_L}^c \equiv (N_{1_L})^c = (1/2) (1+\gamma_5) i \gamma^2 N_1^*$.
The interactions of $N_1$ with $W'$ take the following form:
\be
\frac{g_{_{\rm R}} } {\sqrt 2}  \, W'_\nu \, \left(  \overline N_{1_R} \gamma^\nu e_R +  \overline N_{1_L}^c \gamma^\nu \tau_R
 \right)  + {\rm H.c.}  
\label{eq:coupling}
\ee
The interaction of $N_R^\mu$ remains as in Eq.~(\ref{eq:WpInt}).

We will assume in what follows that the gauge and mass eigenstates of the charged leptons are identical. In Section 4 we will
show that this can be a consequence of the same global $U(1)$ symmetry responsible for the flavor structure in Eq.~(\ref{eq:WpInt}).

\bigskip

\section{Predictions for the LHC}\setcounter{equation}{0}

The couplings of the 4-component $N_1$ fermion (formed of the 2-component fermions $N_R^e$ and $N_R^\tau$) 
to the $W^\prime$ boson
displayed in Eq.~(\ref{eq:coupling}) are peculiar: the $\overline N_1$ antifermion interacts with the electron, while the $N_1$ fermion 
interacts with the $\tau$. This has profound implications for the LHC phenomenology. 

For the mass ordering of interest here, $m_{N_1} < M_{W'} \lesssim m_{N^\mu}$, 
the $W^{\prime +} $ boson may undergo leptonic decays only into $e^+ N_1$ or $\tau^+ \overline N_1$,  with equal branching fractions. Similarly, the only leptonic decay channels of the $W^{\prime -} $ boson are  $e^- \overline N_1$ and $\tau^- N_1$. 

The $N_1$ fermion predominantly decays via an off-shell $W'$ into a quark-antiquark  pair and an $e^-$ or a  $\tau^+$, again with equal branching fractions.
 Note that  $N_1$ decays into a quark-antiquark pair and an $e^+$ or a  $\tau^-$ are forbidden by the $U(1)$ flavor symmetry introduced in Section 2,
 while the $\overline N_1$ antifermion decays into these final states are allowed.
Consequently, the following cascade decays have equal branching fractions:   
\bear  
&&  \hspace*{-1cm}
B( W^{\prime +} \! \!\to e^+ N_1 \to e^+ e^- jj ) = B( W^{\prime +} \!\! \to \tau^+ \overline N_1 \to \tau^+ \tau^- jj ) 
\nonumber \\ [2mm]
&& \hspace*{3.9cm}
= B( W^{\prime +} \! \!\to e^+ N_1 \to e^+ \tau^+ jj ) 
\nonumber \\ [2mm]
&& \hspace*{3.9cm} = B( W^{\prime +}  \!\!\to \tau^+ \overline N_1 \to \tau^+ e^+ jj )  ~~.
\label{eq:BRs}
\eear
CPT invariance implies that these branching fractions are also equal to those for the cascade decays of $W^{\prime -} $ into $e^+ e^- jj$, $ \tau^+ \tau^- jj$, $e^- \tau^- jj$, 
and $\tau^- e^- jj$. 

The $W'$ widths into same-sign $eejj$ or $\tau\tau jj$, as well as into opposite-sign $e^\pm\tau^\mp jj$ vanish. These are immediate consequences 
of the $W^\prime$ couplings shown in Eq.~(\ref{eq:coupling}) in the Dirac basis.\footnote{The same result can be obtained using 
the Majorana basis. After diagonalizing the mass matrix shown in Eq.~(\ref{MR}), 
the 2-component mass eigenstates are given by $i(N^e_R - N^\tau_R)/\sqrt{2}$ and $(N^e_R + N^\tau_R)/\sqrt{2}$. The 
destructive interference between decay amplitudes proceeding through these two Majorana states gives vanishing widths for 
$W^{\prime \mp} \!\to e^\mp e^\mp jj$. Feynman rules for Majorana fermions useful for this computation are given in \cite{Denner:1992vza,Han:2012vk}.}
Both the same-sign and opposite-sign
$\mu\mu jj$ signals are negligible as long as $m_{N^\mu}$ is not smaller than $M_{W'}$ by more than 5\% or so. 

The observed excess events in the $e^+e^-jj$ final state  \cite{Khachatryan:2014dka}  can  be explained in this theory, together with the other 
hints for a $W'$ near 2 TeV discussed in \cite{Dobrescu:2015qna} for $g_{_{\rm R}} \approx 0.5$. 
Note first that compared to the case where the Dirac partner of $N_R^e$ is a new fermion,
the $e^+e^-jj$ rate predicted in the theory discussed here has one less parameter because the  $N_1$ coupling to an electron and a $W'$ is fixed
in Eq.~(\ref{eq:coupling}).   
The ensuing branching fraction of $N_1$ into an electron and a quark-antiquark pair is larger 
by a factor of 2 than in the baseline model analyzed in \cite{Dobrescu:2015qna}, 
while the branching fraction of $W' \! \to e N_1$ is larger by a factor of  4.
As these changes increase the $e^+e^-jj$ rate (by a factor of 8), they can easily be compensated by the phase-space suppression of the $W' \! \to e N_1$ width.
We find that the predicted $e^+e^-jj$ rate is consistent with the CMS excess when the $N_1$ mass  satisfies 1.4~TeV~$\lesssim m_{N_1} \lesssim$~1.7~TeV for 
$M_{W'} \approx 1.9$ TeV.

As a result of the equalities between branching fractions, there are a few striking predictions for the LHC. First the cross section for 
$pp \to W'\! \to e^+ e^- jj $ is equal to that for $pp \to W'\! \to \tau^+ \tau^- jj $.
A comparison of the same-sign $e\tau jj$ cross sections with the opposite sign $eejj$ cross section is less straightforward because 
the production cross section in $pp$ collisions for $W^{\prime +}$ is larger than that for $W^{\prime -}$. There is, however, a simple relation:
the $pp \to W' \! \to e^+ e^- jj $ cross section is exactly one half of the sum of 
the $pp \to W^{\prime +} \! \to e^+ \tau^+ jj $ and $pp \to W^{\prime -} \! \to e^- \tau^- jj $ cross sections.
Thus,
\bear
&& \hspace*{-1.9cm} \sigma (pp \to W' \! \to e^+ e^- jj ) = \sigma (pp \to W' \!\to \tau^+ \tau^- jj ) 
\nonumber \\ [2mm]
&& \hspace*{2.4cm}
= \frac{1}{2} \left[ \sigma (pp \to W' \!\to e^+ \tau^+ jj ) + \sigma (pp \to W' \!\to e^- \tau^- jj) \rule{0pt}{14pt} \right] ~~.
\eear

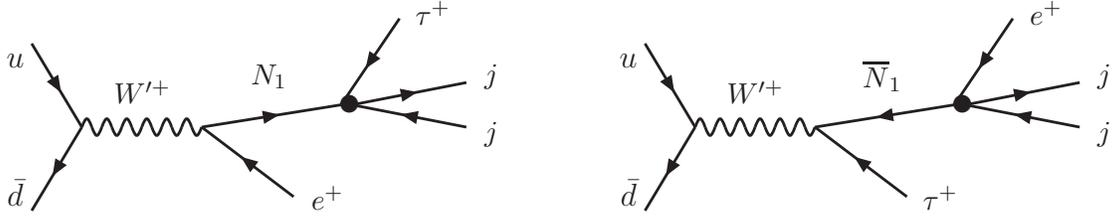
\begin{figure}[t]
\begin{center} 
{
\unitlength=2. pt
\SetScale{1.05}
\SetWidth{1.}      
\normalsize    
{} \allowbreak
\begin{picture}(100,88)(4,-37)
\ArrowLine(4,80)(22,50)
\ArrowLine(22,50)(4,20)
\Photon(22,50)(65,50){3}{6}
\ArrowLine(65,50)(115,58) \ArrowLine(136,89)(115,58)  \ArrowLine(160,50)(119,58) \ArrowLine(119,58)(160,66)  
\ArrowLine(98,25)(65,50) 
\Text(-1,39)[c]{$u$}\Text(-1,14)[c]{$\bar d$}\Text(23,33)[c]{\small $ W^{\prime +}$}
\Text(47,36)[c]{\small $N_1$}\Text(58,13)[c]{\small $e^+$} \Text(61.2,30.5)[c]{ \Large $\bullet$}
\Text(78,48)[c]{\small $\tau^+$} \Text(89,36)[c]{\small $j$}\Text(89,25)[c]{\small $j$}
\end{picture}
\begin{picture}(100,88)(-10,-37)
\ArrowLine(4,80)(22,50)
\ArrowLine(22,50)(4,20)
\Photon(22,50)(65,50){3}{6}
\ArrowLine(115,58)(65,50) \ArrowLine(136,89)(115,58)  \ArrowLine(160,50)(119,58) \ArrowLine(119,58)(160,66)  
\ArrowLine(98,25)(65,50)
\Text(-1,39)[c]{$u$}\Text(-1,14)[c]{$\bar d$}\Text(23,33)[c]{\small $ W^{\prime +}$}
\Text(47,36)[c]{\small $\overline N_1$}\Text(58,13)[c]{\small $\tau^+$} \Text(61.2,30.5)[c]{ \Large $\bullet$}
\Text(78,48)[c]{\small $e^+$} \Text(89,36)[c]{\small $j$}\Text(89,25)[c]{\small $j$}
\end{picture}
}
\end{center}
\vspace*{-3.4cm}
\caption{Processes for resonant $e^+\tau^+ jj$ production at the LHC. The Dirac fermion $N_1$, formed of the $N_R^e$ right-handed neutrino 
and the charge-conjugate of $N_R^\tau$, has interactions with $\tau$ that violate fermion number (note the arrows) 
according to  Eq.~(\ref{eq:coupling}). The $\bullet$ represents the 4-fermion interaction mediated by an off-shell $W^{\prime -}$.
   }
\label{fig:ZN}
\end{figure}

These predictions can be tested in various ways in Run 2 of the LHC. The production cross section for a $W'$ of mass near 2 TeV 
grows by a factor of 5 at $\sqrt{s} = 13$ TeV compared to $\sqrt{s} = 8$ TeV. Thus, the approximately ten $eejj$ signal events observed by CMS 
with 20 fb$^{-1}$ in Run 1 imply about 500 $e^\pm \tau^\pm jj$ signal events (see Figure 1) 
with 100 fb$^{-1}$ in Run 2, if the efficiency of the event selection is not modified. 
Even though the backgrounds increase in Run 2, that number of events would allow the observation of same-sign $e\tau jj$ signals independently with leptonic or hadronic $\tau$ decays.

So far we focused on $N_1$ decays into a lepton and two light quarks, which have the dominant branching fractions. Other 
$N_1$ decay modes potentially observable in Run 2 include a lepton plus $t\bar b$ or $WZ$, which occur via an off-shell $W'$,
as well as a lepton plus a $W$ boson. All these decays lead to the same cross-section relations as those involving light quarks. 
For example,  $\sigma (pp \to e^+ \tau^+ \bar t \, b ) + \sigma (pp \to e^- \tau^- \bar t \, b ) = 2  \sigma (pp \to e^+ e^- \bar t \, b )$.

The $SU(2)_L \times SU(2)_R \times U(1)_{B-L}$ gauge group implies that besides the $W'$ boson there is a  $Z'$ boson. 
The breaking of $SU(2)_R \times U(1)_{B-L}$ by an  $SU(2)_R$-triplet VEV, as assumed here, implies that the $Z'$ is substantially heavier 
than the $W'$ boson. For $M_{W'} \approx 1.9$ TeV and $g_{_{\rm R}}$ in the 0.45--0.6 range,  the $Z'$ mass satisfies 
3.4~TeV~$< M_{Z'} <$~4.5~TeV \cite{Dobrescu:2015qna}, a range that will be probed in Run 2 of the LHC. 
Our right-handed neutrino sector leads to testable predictions. The $Z' \to \bar N_1 N_1$ decay followed by $N_1$ decays to an electron or $\tau$ and two jets
leads to final states with two leptons and four jets. The flavor structure of the model implies
\be
B(Z' \! \to e^- \tau^- \!\! \, +4j) = B(Z' \! \to e^+ \tau^+ \! \! \,+4j) = B(Z' \! \to e^+ e^- \!\! \, +4j) 
= B (Z' \!\to \tau^+ \tau^- \!+4j) ~.
\ee
An additional prediction of our model is that the same-sign same-flavor channels $Z'\to \tau^\pm \tau^\pm  \! + 4j$ and $ Z' \to e^\pm e^\pm  \! +4j $ are forbidden.

\bigskip\bigskip

\section{Masses for Standard Model neutrinos}\setcounter{equation}{0}

Both the charged lepton masses and Dirac neutrino masses may be generated, in principle, by the following $SU(2)_L\times SU(2)_R \times U(1)_{B-L}$-invariant 
Yukawa couplings to the bidoublet $\Sigma$ and its charge-conjugate state $\widetilde \Sigma$: 
\be
- \overline L_L^\alpha \left( y_{\alpha\beta} \,  \Sigma + \tilde y_{\alpha\beta}  \, \widetilde \Sigma \right) L_R^\beta  + {\rm H.c.} 
\label{eq:sigma}
\ee
where $\alpha,\beta $ are flavor indices,  $L_L^\alpha=(\nu_L^\alpha , \ell_L^\alpha)^\top$ are the $SU(2)_L$
lepton doublets, and $y$ and $\tilde y$ are dimensionless coefficients.
However, depending on the charges of $\Sigma$ and $L_L^\alpha$
under the $U(1)$ global symmetry discussed in Section 2,
the terms shown in Eq.~(\ref{eq:sigma}) may be forbidden. 

The Dirac masses that link the SM neutrinos to the right-handed neutrinos cannot be larger than the MeV scale 
if the right-handed neutrino masses are at the TeV scale (otherwise the SM neutrino masses, obtained from the seesaw mechanism, are too large).
Since the nonzero components of the  bidoublet VEV are of the same order ({\it i.e.}, $\tan\beta \sim 1$ as discussed in Section 2), a fine-tuned cancellation between 
the $y$ and $\tilde y$ coefficients would be needed to keep these Dirac masses much smaller than the
$\tau$ mass. We therefore 
choose global $U(1)$ charges that forbid the interactions in Eq.~(\ref{eq:sigma}); for example, $L_L^e$, $L_L^\mu$ and $L_L^\tau$ have 
same charges as the corresponding right-handed leptons ($-1$, 0, $+1$),
while $\Sigma$ has charge +3.

Let us present a different mechanism for generating the SM lepton masses.
We introduce a gauge-singlet scalar $\phi$ that carries $U(1)$ charge +3, and acquires a VEV $\langle \phi \rangle$.
Consider the following gauge-invariant operators:
\be
- \frac{\tilde  c_{\alpha } }{m_f^3} \; \phi \, \overline L_L^\alpha \; \widetilde \Sigma \,
 {\widetilde T^\dagger \widetilde T}  \, L_R^\alpha   \;  + {\rm H.c.}  ~~,
 \label{eq:sigmaTTtilde}
\ee
where $\tilde  c_{\alpha }$ ($\alpha = e,\mu,\tau$) are dimensionless couplings and $m_f$ is the mass of some heavy fields, which have been integrated out.
These operators generate the charged lepton masses (similarly to the down-type quark masses \cite{Dobrescu:2015yba}) 
once the scalars acquire VEVs:
\be 
(m_e, m_\mu, m_\tau) = v_H \cos\!\beta \; \frac{\langle \phi \rangle u_T^2}{m_f^3} \; (\tilde c_e,  \tilde c_\mu , \tilde c_\tau) ~~.
\label{eq:ml}
\ee
This ensures that the gauge and mass eigenstates of the charged leptons coincide, as mentioned in Section 2.

The SM neutrinos get Majorana masses from the following gauge-invariant dimension-6 operators:
\be
\frac{\eta_{\alpha\beta}}{M^2} \; (L_L^\alpha)^c \; \Sigma \, T \,\Sigma^\dagger \; L_L^\beta ~~.
\label{eq:LLHH}
\ee
The global $U(1)$ allows the $\eta_{\mu\mu}$, $\eta_{e\tau}$ and $\eta_{\tau e}$ coefficients to be nonzero.
Dirac masses between the SM and right-handed neutrinos may also be generated by 
\be
- \frac{C_{\alpha\beta}}{m_f^3} \; \phi \, \overline L_L^\alpha \, \widetilde\Sigma \,
 {T^\dagger T}  \,  L_R^\beta  + {\rm H.c.} ~~,
 \label{eq:sigmaTT}
\ee
 with the result
\be 
m_{D} =   v_H \sin\!\beta \; \frac{\langle \phi \rangle u_T^2}{m_f^3} \; C ~~~.
\ee
Here we ignored the complex phase from the $\Sigma$ VEV, and we collected the  $C_{\alpha\beta}$  coefficients in a $3\times 3$ matrix $C$.
The mass matrices for the charged and neutral leptons are independent of one another and no cancellation is necessary. 
The full $6\times 6$ mass matrix in the neutrino sector has the following block structure:
\be 
\mathcal{M}_\nu = \left( 
\ba{cc}   
m_L & m_{D} \\ [2mm]
m_{D}^\top & M_R
\ea \right)  ~~,
\label{eq:massmatrix}   
\ee
where the Majorana mass matrix for the right-handed neutrinos, $M_R$, is shown in Eq.~(\ref{MR}), and the Majorana mass matrix for the SM neutrinos, $m_L$,
arises from operators (\ref{eq:LLHH}).

All Lagrangian terms discussed above are invariant under  the global $U(1)$.
However, the spontaneous breaking of this symmetry implies the existence of a Nambu-Goldstone boson $\theta_\phi$  that couples to leptons. In order to satisfy phenomenological 
constraints, the global $U(1)$ must also be explicitly broken so that $\theta_\phi$  becomes heavy (alternatively, the global symmetry is a $Z_n$ group, avoiding the
presence of  Nambu-Goldstone bosons).
Note that $\langle \phi \rangle$ may be much larger than the $SU(2)_R$ breaking scale, so that even small explicit $U(1)$ breaking terms 
may push the mass of $\theta_\phi$ above the reach of the LHC.

Lagrangian terms that explicitly break the global $U(1)$ may also have important contributions
to the elements of the $m_{D}$ and $m_L$ matrices. The smallness of these mass terms implies negligible effects in
$M_R$. 
Thus, there is enough freedom to accommodate the masses and mixings of the active neutrinos. 
After integrating out the right-handed neutrinos, whose masses are at the TeV scale, the SM neutrinos acquire Majorana masses.

Let us comment on the particular case where $m_L$ is negligibly small.
Given the particular structure of $M_R$, the minimal possibility for $m_{D}$ that 
reproduces the light neutrino squared-mass differences and mixing angles observed at neutrino oscillation experiments~\cite{Gonzalez-Garcia:2014bfa} is the following:
\be
m_{D} = \left( \ba{ccc}   m_{11} & 0 & m_{13} \\ m_{21} & 0 & m_{23}  \\   m_{31} & 0 & m_{33}  \ea \right) \equiv
\left(   m_{D1} \; , \; 0 \; , \; m_{D3} \right)  ~~~.
\label{eq:mDirac}
\ee
The Majorana mass matrix for the light active neutrinos is  obtained by a TeV-scale seesaw:
\be
m_{\nu}=  m_{D1} \frac{1}{m_{N1}} m_{D3}^{\top}.
\label{mnu}
\ee
Notice that in this minimal model only the Dirac fermion, formed by $N_R^e$ and $(N_R^\tau)^c$, participates
in the light neutrino mass generation,  because the Majorana state $N_R^\mu$ decouples.

The neutrino mass matrix given in Eq.~(\ref{mnu}) has a zero mode and thus one of the light neutrino masses vanishes, while the other two generate 
the solar and atmospheric neutrino mass differences. This scenario resembles 
the minimal linear seesaw~\cite{Malinsky:2005bi} studied in detail in \cite{Gavela:2009cd}, where it was
shown that the flavor structure of the neutrino Dirac mass terms, $m_{D1}$ and $m_{D3}$, is completely fixed by 
neutrino oscillation data up to a global factor. 

In order to generate a third light neutrino mass (and have more freedom in the parameter space), 
the second column of $m_{D}$ should be switched on. If this is the case,  $N_R^\mu$ would also participate in the 
generation of light neutrino masses.

Our TeV-scale right-handed neutrino sector is not currently constrained by
low-energy observables such as neutrinoless double beta decay or lepton-flavor violating decays. 
Generically, the rate for neutrinoless double beta decay  mediated by the $W'$ and the right-handed neutrinos is
tightly correlated to the  $p p \rightarrow e^{\mp} e^{\mp} jj $ cross section. 
Our mechanism, which forbids the same-sign $e e j j$ signal at the LHC, also forbids neutrinoless double beta decay mediated by two $W'$ bosons. 

There is, however, a new physics contribution to the neutrinoless double beta decay mediated by two $W$ bosons. 
This arises from one insertion of the mixing between a SM neutrino and a right-handed 
neutrino [see $m_{13}$ in Eq.~(\ref{eq:mDirac})], and one insertion of the mixing between the $SU(2)_L\times SU(2)_R$ gauge bosons. 
The latter has an upper limit  \cite{Dobrescu:2015yba} of $\sin\theta_+ \leq (g_{_{\rm R}}/g) (M_W/M_W')^2 \approx 10^{-3}$, where $g$ is the SM weak coupling.
The contribution of this type \cite{Nemevsek:2012iq} to the effective neutrino mass relevant for neutrinoless double beta decay is given by 
\bear
m_{\beta\beta} & \simeq &  \frac{g_{_{\rm R}}}{g}   \sin\theta_+  \; \frac{m_{13}}{m_{N_1}} \langle p \rangle 
\nonumber \\ [2mm]
& \approx &  0.1 \, 
\text{eV} \; \left(\frac{ \sin\theta_+}{10^{-3}}\right)
\left(\frac{m_{13}}{1\,  \text{MeV}}\right) \left(\frac{1 \, \text{TeV}}{m_{N_1}} \right) \left( \frac{\langle p \rangle}{100 \text{ MeV}}\right)  ~~,
\eear
where $\langle p \rangle$ is the momentum transfer in the nuclear transition.
For $m_{13}\sim 0.3$ MeV and $m_{N_1}\approx 1.5$ TeV, this would be within the 
reach of future neutrinoless double beta decay experiments. 
 In addition, there is the usual light neutrino contribution with two $W$ propagators. 
 This  is suppressed if the light neutrino spectrum has normal mass ordering, so a 
 neutrinoless double beta decay signal would provide information about  $N_1$. 
 For an inverted ordering  (or normal ordering with a quasi-degenerate spectrum), the contributions from  $N_1$ and from the light neutrinos are comparable. Thus, the total rate could be slightly enhanced or even suppressed with respect to the case of only  light neutrinos.

Due to the large $e-\tau$ flavor mixing in the right-handed neutrino sector in our model, it is relevant 
to ask whether there are  constraints from charged lepton-flavor violating processes. 
The $\tau \to e \gamma$ contribution from loops involving the $W'$ and right-handed neutrinos is suppressed because 
 the flavor structure of our model imposes that lepton-flavor violation and lepton-number violation can only occur simultaneously. 
The main contribution to charged lepton-flavor violating decays in our model is in the $ \tau^\mp \! \to \mu^\mp \mu^\mp e^\pm$ channel. This decay proceeds 
via a box diagram with two $W'$ bosons, where one of the fermion lines involves only muon-flavor states, while the $\tau-e$ flavor transition takes place along the 
second fermion line via an $N_1$ exchange. The branching fraction for this process is of the order of $g_{_{\rm R}}^8 M_W^4/(4 \pi  g M_{W'})^4 \approx 10^{-12}$, where we have taken into 
account that the $m_{N^\mu}/M_{W'}$ and  $m_{N_1}/M_{W'}$ ratios are of order one. 
This prediction is below the current experimental limit $B(\tau^- \! \to e^+ \mu^- \mu^-) < 1.7\times 10^{-8}$ 
\cite{Agashe:2014kda}. 

As the muon-lepton number is violated in Eq.~(\ref{eq:flavorYuk}) by two units, 
the $U(1)$-invariant contributions to $\mu \to e e e $, $ \mu \to e $ conversion in nuclei, or  $ \mu \to e \gamma $ 
vanish. The explicit breaking of the $U(1)$ flavor symmetry leads to contributions through the mixing of light and heavy neutrinos. 
Given that in our model flavor-violating processes occur at one loop, the most sensitive probe is $ \mu \to e \gamma $.
The predicted branching fraction, though, is below the current limit \cite{Adam:2013mnn}
because it is suppressed by $(\alpha/\pi) \sin\!^2\theta_+ \lesssim 10^{-9}$ as well as  $(m_{21}/m_\mu)^2 \lesssim 10^{-5}$. 

\bigskip\bigskip

\section{Conclusions}

In a SM extension  with three right-handed neutrinos of masses at the TeV scale, 
we have proposed a flavor structure that pairs the right-handed electron-neutrino and 
the right-handed tau-neutrino, forming a Dirac fermion ($N_1$) whose mass violates lepton number. 
The right-handed muon-neutrino ($N^\mu_R$) remains a purely Majorana state. 
This flavor structure is enforced by a global $U(1)$ symmetry, and leads to peculiar  
interactions [see Eq.~(\ref{eq:coupling})] of 
the $W'$ boson associated with the $SU(2)_L\times SU(2)_R \times U(1)_{B-L}$ gauge group.

Our flavor symmetry predicts specific relations between the decay widths of the TeV-scale neutrinos.
$N_1$ decays into an $e^-$ or a $\tau^+$ and a $W'$ (which is off-shell for $m_{N_1} < M_{W'}$) 
or a $W$ (through $W_L-W_R$ mixing). 
The branching fractions for decays involving $e^-$ are equal to those involving $\tau^+$. 
The $N^\mu_R$ Majorana fermion decays, also with 50\% branching fractions,
into a $\mu^-$ or a $\mu^+$ and a $W'$ (which is on-shell assuming $m_{N^\mu} > M_{W'}$).
 
Furthermore, the decays of the $W'$ into a SM lepton and a heavy neutrino have tightly correlated branching fractions,
as shown in Eq.~(\ref{eq:BRs}). 
As a consequence, the decay of the $W'$ boson into an electron and a right-handed neutrino
produces an $e^+e^-jj$ signal, while the rate for same-sign $e^\mp e^\mp jj$ events automatically vanishes.
This provides a compelling explanation for the excess of opposite-sign $e^+e^-jj$ events 
with an invariant mass near 2 TeV
reported by the CMS Collaboration \cite{Khachatryan:2014dka}.

The flavor symmetry leads to additional predictions at the LHC. First, 
the opposite-flavor same-sign processes $pp \to W^{\prime +} \! \to \tau^+ e^+  jj$ 
and $pp \to W^{\prime -} \! \to \tau^- e^- jj$  are allowed; the sum or their rates is exactly twice
the rate for the same-flavor opposite-sign process $pp \to W' \!  \to e^+ e^-  jj$.
Second, the rates for the two same-flavor opposite-sign processes, $W' \! \to e^+ e^-  jj$ and 
$W' \! \to \tau^+ \tau^-  jj$, are equal.
These and other predictions discussed in Section 3 can be tested in Run 2 of the LHC.

The masses for active neutrinos get contributions from both a TeV-scale seesaw mechanism and 
lepton-number violating dimension-6 operators involving only left-handed fields.
Ignoring the latter, there are Dirac mass terms at the MeV scale that link the left- and right-handed neutrinos; 
after integrating out the right-handed neutrinos at the TeV scale, the left-handed neutrinos acquire Majorana masses
at the sub-eV scale.
The minimal choice for the Dirac mass matrix that reproduces the neutrino oscillation data leaves one of the light neutrinos massless. 
In this scenario, the light neutrinos acquire Majorana masses only from their Yukawa interactions with $N_1$.
A more general choice for the Dirac mass matrix 
would allow the third left-handed neutrino to acquire mass. In that case, the right-handed muon-neutrino would also be involved in the generation of active 
neutrino masses.

\bigskip \bigskip  \bigskip

{
{\bf Acknowledgments:} \ 
We would like to thank Frank Deppisch, Janusz Gluza, Tomasz Jelinski,  Zhen Liu, and especially Patrick Fox for stimulating communications. 
JLP would like to thank Fermilab for hospitality and partial support during the completion of this work. JLP and PC acknowledge
financial support by the European Union through the ITN INVISIBLES (Marie Curie Actions, PITN-GA-2011-289442-
INVISIBLES). PC would like to thank the Mainz Institute for Theoretical Physics for hospitality and partial support during the completion of this work. 
Fermilab is operated by the Fermi Research Alliance
under contract DE-AC02-07CH11359 with the U.S. Department of Energy. JLP was
partially supported by grant 2012CPPYP7 (Theoretical Astroparticle Physics) under the program PRIN 2012
funded by MIUR and the INFN program on Theoretical Astroparticle Physics (TASP). 
}

\vfil
\end{document}